\documentstyle[epsfig]{aipproc}

\newcommand{\eq}[1]{Eq.~(\ref{#1})}

\def\beq{\begin{equation}}
\def\eeq{\end{equation}}
\def\beqa{\begin{eqnarray}}
\def\eeqa{\end{eqnarray}}

\newcommand{\EQ}{\begin{equation}}
\newcommand{\EN}{\end{equation}}
\newcommand{\bea}{\begin{eqnarray}}
\newcommand{\ena}{\end{eqnarray}}

\renewcommand{\a}{\alpha}

\newcommand{\ve}{\varepsilon}

\begin{document}

\title{Two-loop gluon diagrams from string theory}

\author{Lorenzo Magnea$^*$ and Rodolfo Russo$^{\dagger}$}

\address{NORDITA\thanks{On leave from Universit\`a di Torino, 
Italy.}\\
Blegdamsvej 17, DK--2100 Copenhagen \O, Denmark\\
Dipartimento di Fisica, Politecnico di Torino
\thanks{The work presented here is part of an ongoing collaboration with
P. Di Vecchia, A. Lerda and R. Marotta}\\
Corso Duca degli Abruzzi 24, I--10129 Torino, Italy\\}

\maketitle

\begin{abstract}
We briefly review the string technology needed to calculate Yang-Mills
amplitudes at two loops, and we apply it to the evaluation
of two-loop vacuum diagrams.

\end{abstract}

\section*{Introduction}

It is well known~\cite{lm:all} that string theory is a powerful
calculational tool for the evaluation of tree and one-loop amplitudes
in Yang-Mills theory and gravity. During the past two years considerable
progress has been made towards the extension of string-inspired techniques
to more than one loop~\cite{lm:us}. The calculation of the two-loop 
Yang-Mills vacuum diagrams, presented here, is the first simple 
application of this formalism to gauge theories beyond one loop.

The general features of string-inspired techniques remain unchanged
to all orders in perturbation theory: the field theory limit is
obtained by taking the string tension $T = 1/(2\pi\alpha')$ to infinity,
decoupling all massive string modes; the corners of string moduli space 
contributing to the field theory result are those where the integrand 
of the string amplitude exhibits a singular behavior, and in these regions 
string moduli are naturally related to Schwinger proper times in field 
theory; finally, string-derived amplitudes are written in a form which
is strongly reminiscent of the world-line formalism in field 
theory~\cite{lm:sss}.

A special feature of calculations done using the bosonic string
is the existence of contact terms, corresponding to four-point vertices
in field theory, that arise as finite remainders of tachyon exchange.
These contributions are present to all orders, however at one loop
they can be eliminated by performing a partial integration at the string 
level. This simplifies calculations, but it obscures the connection
between string-derived rules and field theory. At two loops, we find
that the same prescription that was used at one loop to handle
tachyon exchange (a $\zeta$-function regularization) is sufficient
to obtain the correct result.

It should be noted that the string derivation is not tied to any special
choice of regularization scheme for ultraviolet and infrared divergences. 
While dimensional regularization is naturally implemented and useful for 
practical calculations, any regularization scheme that can be applied at 
the level of Schwinger parameter integrals is just as natural. Thus it is
meaningful to calculate vacuum diagrams, although they vanish in dimensional
regularization.

\section*{Multiloop gluon amplitudes}

Let us begin by recalling the general expression for the color-ordered 
$h$-loop $M$-gluon planar amplitude in the open bosonic 
string~\cite{lm:us}, 
\beqa 
A^{(h)}_P & = & C_h {\cal N}_0^M \int [dm]_h 
\frac{\prod_{i=1}^M dz_i}{dV_{abc}} \prod_{i<j} 
\left[\frac{\exp\left({\cal G}^{(h)}(z_i,z_j)\right)}{
\sqrt{V'_i(0)\,V'_j(0)}}\right]^{2\a' p_i\cdot p_j} 
\label{lm:hmast} \\
& \times & \exp \left[\sum_{i \not= j} \left(
\sqrt{2\a'} p_j\cdot\ve_i 
\,\partial_{z_i} {\cal G}^{(h)}(z_i,z_j) 
+ \, {1\over 2}\ve_i\cdot\ve_j
\,\partial_{z_i}\partial_{z_j}
{\cal G}^{(h)}(z_i,z_j)\right)\right]~, 
\nonumber
\eeqa
where only terms linear in each polarization should be kept, and we omitted
the color factor $N^h\,{\rm Tr}(\lambda^{a_1}\cdots \lambda^{a_M})$.
The dimensionless string coupling constant $g_s$ is related to the 
$d$-dimensional gauge coupling $g_d = g \mu^{(4 - d)/2}$ by
\beq
g_s = \frac{g_d}{2}\,(2\a')^{1-d/4}~.
\label{lm:gs}
\eeq
The fundamental ingredients of \eq{lm:hmast} are the bosonic Green function 
${\cal G}^{(h)}(z_i,z_j)$ (the correlator of two scalar fields on the
$h$-loop string world sheet), and the measure of integration on moduli space
$[dm]_h$. Both these quantities depend only on the genus $h$ of the 
surface and thus represent the building blocks for the calculation of 
all diagrams at $h$ loops with an arbitrary number of external states.
In particular, the string Green function acts as a generator of the
specific world-line Green functions found for particle diagrams of 
different topology, to which it reduces in the appropriate corners of 
moduli space. 
Explicit expressions for the Green function, for the projective
transformations $V_i(z)$, which define local coordinate systems around 
the punctures, and for the normalization constants $C_h$ and ${\cal N}_0$
can be found in Ref.~\cite{lm:us}; here we want to focus on the 
measure of integration, since it encodes all the information needed for
the derivation of the vacuum diagrams. It is given by
\begin{eqnarray}
[dm]_h & = & 
\prod_{\mu=1}^{h} \left[ \frac{dk_\mu d \xi_\mu d \eta_\mu}{k_\mu^2
(\xi_\mu - \eta_\mu)^2} ( 1- k_\mu )^2 \right]   \label{lm:meas} 
\left[\det \left( - i \tau_{\mu \nu} \right) \right]^{-d/2} 
\\ & \times &
\prod_{\alpha}\;' \left[ \prod_{n=1}^{\infty} ( 1 - k_{\alpha}^{n})^{-d}
\prod_{n=2}^{\infty} ( 1 - k_{\alpha}^{n})^{2} \right]~.   
\nonumber
\end{eqnarray}
The various ingredients of this formula have a geometric intrepretation
on a Riemann surface of genus $h$. In particular, $\tau_{\mu \nu}$ is the 
period matrix, while $k_{\mu}$, $\xi_{\mu}$ and $\eta_{\mu}$ are 
the moduli of the surface in the Schottky parametrization~\cite{lm:cop}; 
the primed product over $\alpha$ denotes a product over conjugacy classes 
of elements of the Schottky group, where only elements that cannot be 
written as powers of other elements must be included. Three of $2 h + M$
parameters $\xi_\mu$, $\eta_\mu$ and $z_i$ can be fixed using an overall 
projective invariance of the amplitude. The fixing of this invariance
introduces the projective invariant volume element $dV_{abc}$ in 
\eq{lm:hmast}. Including the ``multipliers'' $k_\mu$, one is left 
with $3 h - 3 + M$ variables, the correct number of independent moduli 
for a Riemann surface of genus $h$ with $M$ punctures. In the field theory
limit, only the region in moduli space in which the multipliers $k_\mu$
are small gives finite contributions.

\section*{Yang-Mills vacuum diagrams}

Now we turn to the study of vacuum diagrams at two loops, specializing
the formulas of the previous section to the case $M=0$, $h=2$. First, we
use projective invariance to fix $\xi_1$, $\xi_2$ and $\eta_1$
to $\infty$, $1$ and $0$ respectively. Then we evalute \eq{lm:meas} in the 
field theory limit. To this end, we expand it in power of $k_{\mu}$, and  
we ignore all terms that are quadratic in one multiplier, since 
they correspond to the undesired exchange of a massive spin 2 state. 
In this approximation, \eq{lm:hmast} becomes 
\begin{eqnarray} 
\label{lm:exp1}
A^0_2 & = & N^3 \, C_2 \, (2\pi)^d \int {dk_1\over k_1^2} {dk_2\over 
k_2^2} {d\eta_2\over (1-\eta_2)^2} \;  
\\ \nonumber &\times &
\left[1+(d-2)(k_1+k_2)+\left((d-2)^2+d\,(1-\eta_2)^2{1+\eta_2^
2 \over \eta_2^2}\right)k_1 k_2\right]
\\ \nonumber &\times &
\Bigg[\ln k_1\ln k_2 - \ln^2\eta_2 + {2\,(1-\eta_2)^2\over 
\eta_2} (k_1\ln k_1 + k_2\ln k_2) +
\\ \nonumber & &
+ {4\,(1-\eta_2)^4\over \eta_2^2}\left(1 + {1+\eta_2 \over 1- \eta_2}
\ln\eta_2 \right) k_1 k_2\Bigg]^{-d/2}~~~. 
\end{eqnarray}
The region of integration can be deduced by studying the Schottky 
representation of the two--annulus~\cite{lm:scal}. In the small 
$k$ limit, it is given by $0\leq\sqrt{k_2}\leq \sqrt{k_1}\leq\eta_2\leq 1$.
The fixed point $\eta_2$ can be interpreted as the distance between the two 
loops, so that we can tentatively identify the region $\eta_2\rightarrow 
1$ as related to reducible diagrams, and the region $\eta_2\rightarrow 
0$ as related to irreducible ones.

In the region $\eta_2\rightarrow 1$, in order to isolate the contribution
of massless states, we must extract from the integrand of \eq{lm:exp1} 
the term proportional to $k_1^{-1} k_2^{-1} (1-\eta_2)^{-1}$.
Then the field theory results are recovered if one introduces the
Schwinger proper times $t_i=\a'\ln k_{i}$, $t_3=\a' \ln(1-\eta_2)$, which 
must remain finite as $\a'\rightarrow 0$. It can be checked that no terms
survive in the limit $\a'\rightarrow 0$, which is the stringy way to
say that the reducible diagram we are considering is zero. There is
however a contact interaction leftover from tachyon exchange in the limit
$\eta_2\rightarrow 1$. This is obtained by isolating the term in 
\eq{lm:exp1} that is proportional to $k_1^{-1} k_2^{-1}(1-\eta_2)^{-2}$,
and requiring that the integrand be independent of $\eta_2$ except for the
tachyon double pole. The double pole is then regularized using a
$\zeta$-function regularization, as
\beq
\label{lm:reg}
\int_0 \frac{d x}{x^2} \sim \int_0 d x \sum_{n=1}^\infty n {\rm e}^{- n x}
\sim \sum_{n=1}^\infty 1 \sim \zeta(0) = -{1\over 2}~~.
\eeq
The field theory contribution to the remaining integral over the proper
times $t_1$ and $t_2$ is given by the term that does not depend on $\a'$,
\beq
\label{lm:tach}
A^0_2\big|_{\eta_2\rightarrow 1} = - \frac{g_d^2}{(4\pi)^d}
N^3 \frac{(d-2)^2}{4} \int\limits_0^\infty dt_1\, dt_2 (t_1 t_2)^{-d/2}~.
\eeq

Let us now turn to the region $\eta_2\rightarrow 0$. Here it is convenient 
to introduce the variables $q_1 = k_2/\eta_2$, $q_2 = k_1/\eta_2$, 
$q_3 = \eta_2$, which are directly related to the field theory proper times 
by $t_i=\a'\ln q_{i}$. The region of integration now takes the form 
$0\leq q_1\leq q_2\leq q_3\leq 1$, and the term that survives in the 
limit $\a'\rightarrow 0$ is given by
\begin{equation} 
\label{lm:asym}
A^0_2\big|_{q_3\rightarrow 0} = \frac{g_d^2}{(4\pi)^d} N^3 d (d-2) 
\int\limits_0^\infty 
dt_2 \int\limits_0^{t_2} dt_1 \int\limits_0^{t_1} dt_3
{t_1+t_2+2t_3 \over (t_1 t_2+t_1 t_3+t_2 t_3)^{1+d/2}}.
\end{equation}
The integrand of the last equation is not symmetric in the proper times 
$t_i$. This prevents us from rewriting the integrations of \eq{lm:asym} as 
indipendent, thus identifying this contribution with that of the 
irreducible field theory diagram with three propagators. However we expect
that also as $\eta_2 \to 0$ there may be contributions with only two
propagators, which might, and do, symmetrize \eq{lm:asym}. In this case
the relevant contribution does not involve the regularization of a tachyon 
pole, but it simply comes from an integration region in which 
two moduli are kept very close to each other. In our case the only
such region can be parametrized as $q_3 = y q_2$, where we do not associate 
any proper time to the variable $y$, which is kept finite.
This region contributes
\beq
\label{lm:asym2}
A^0_2\big|_{q_3\rightarrow q_2} = -
\frac{g_d^2}{(4\pi)^d} N^3 2 (d-2) \int_0^\infty 
dt_2 \int_0^{t_2} dt_1 (2 t_1 t_2+t_2^2)^{-d/2}~,
\eeq
where the integrand can be rewritten as
\beq
\label{lm:id}
(2 t_1 t_2+t_2^2)^{-d/2} = (t_1 t_2)^{-d/2} + \int_0^{t_1} dt_3 
\frac{t_1+t_2}{(t_1 t_2 + t_1 t_3 + t_2 t_3)^{1 + d/2}}~.
\eeq
The $(-1)$ factor in \eq{lm:asym2} comes from the integration 
over $y$, where only the contribution from the lower limit of integration
is relevant in the field theory limit. \eq{lm:asym2} thus 
symmetrizes \eq{lm:asym}, as well as contributing to the contact term. 
It is easy to check that the sum of Eqs.~(\ref{lm:tach}), (\ref{lm:asym}) 
and (\ref{lm:asym2}) correctly reproduces the sum of the corresponding 
Feynman diagrams in Yang-Mills theory. 
In particular, it is amusing to notice that, after doing the $t_3$ 
integral, the sum of \eq{lm:asym} and \eq{lm:asym2} vanishes, so
that the entire result for the two-loops vacuum diagrams actually comes 
from the contact term arising from tachyon exchange, given by 
\eq{lm:tach}.

\section*{Conclusions}

We have reviewed some of the results of string perturbation theory that
lead to an efficient organization of multiloop Yang-Mills amplitudes, and
we have described the simplest application of the method at two loops. The
next natural step is the evaluation of the two-loop two-point function,
which is expected to contain the two-loop Yang-Mills $\beta$ function,
since string amplitudes lead to background field gauges. Work in this 
direction is in progress.


\begin{references}
\bibitem{lm:all} See, for example, M.L. Mangano and S.J. Parke,
{\it Phys. Rep.} {\bf 200} (1991) 301; and
Z. Bern, L. Dixon and D.A. Kosower, {\it Ann. Rev. Nucl. Part. Sci.}
{\bf 46} (1996) 109, {\tt hep-ph/9602280},  and references therein.
\bibitem{lm:us} P. Di Vecchia, A. Lerda, L. Magnea and R. Marotta,
{\it Phys. Lett.} {\bf B 351} (1995) 445, {\tt hep-th/9502156}.
P. Di Vecchia, A. Lerda, L. Magnea, R. Marotta and R. Russo, 
{\it Nucl. Phys.} {\bf B 469} (1996) 235, {\tt hep-th/9601143};
in Erice, {\it Theor. Phys.} (1995), {\tt  hep-th/9602055};
proceedings of the 29th Ahrenshoop Symposium, Buckow, Germany,
(August 1995), {\tt hep-th/9602056}; proceedings of the Workshop on Gauge 
Theories, Applied Supersymmetry and Quantum Gravity,
London, England, (July 1996), {\tt hep-th 9611023}.
\bibitem{lm:sss} C. Schubert, {\it Acta Phys. Polon.} {\bf B 27} (1996) 
3965, {\tt hep-th/9610108}, and references therein .
\bibitem{lm:cop} P. Di Vecchia, F. Pezzella, 
M. Frau, K. Hornfeck, A. Lerda and S. Sciuto, {\it Nucl. Phys.} {\bf B 322}
(1989) 317.
\bibitem{lm:scal} P. Di Vecchia, A. Lerda, L. Magnea, R. Marotta and 
R. Russo, {\it Phys. Lett.} {\bf B 388} (1996) 65, {\tt hep-th/9607141}; 
K. Roland, {\it Phys. Lett.} {\bf B 289} (1992) 148.
\end{references}
\end{document}